\documentclass{aa}
\usepackage{graphicx} 
\usepackage{epsf} 

\begin{document}
\title{AGAPEROS: Searching for variable stars in the LMC
Bar\thanks{This work is based on data collected by the EROS and DENIS
collaborations.}}
\subtitle{II. Temporal and near-IR analysis of Long-Period Variables}

\author{T.~Lebzelter\inst{1}
\and M.~Schultheis\inst{2} 
\and A.~L.~Melchior\inst{3}
}
\authorrunning{T.~Lebzelter et al.}

\offprints{schulthe@iap.fr} 
\institute{Institut f\"ur Astronomie, T\"urkenschanzstr. 17, A-1180
Wien, Austria 
\and Institut d'Astrophysique de Paris, CNRS, 98bis Bd Arago, 75014
Paris     
\and  LERMA, Observatoire de Paris, 61 av.~de l'Observatoire, 75014
Paris, France}  
  
\voffset 1.0truecm

\date{Received ....  / Accepted .. .........}  
  
\sloppy  
  
\abstract{
We analysed the light curves of a large sample of long period variables
in the LMC from the AGAPEROS catalogue. The (non)regularity of the light
change is discussed in detail showing that the majority of the light
curves cannot be described properly by a single period. We show that
semiregular and small amplitude variability do not necessarily correlate
as has been assumed in several previous studies. Using near-infrared
data from the DENIS survey we correlate the light change with colours and 
luminosities of the objects. These results are used to compare
long period variables in the LMC with LPVs in the Galactic Bulge and
in the solar neighborhood.
} 

\maketitle

  
\section{Introduction} \label{introduction}  
The late stages of stellar evolution are characterized by regular and
irregular light variability, a well-known signature of the stellar
pulsation of Asymptotic Giant Branch stars (hereafter AGB).  These
light changes allow to identify AGB stars over large distances and to
derive the pulsation characteristics (periodicity, etc.), which are
key parameters for understanding the fundamental properties of the
highly extended atmospheres of these stars.  The pulsational
properties have a strong impact on the structure of AGB
stars. Pulsation, as the driving mechanism for the stellar winds,
plays a key role for the high mass loss rates reached during the AGB
phase.

A new era in the study of variable red giants started, when
microlensing surveys produced a large amount of light curves of these
stars, especially for objects in the LMC. The pioneering work by Wood
(\cite{Wood2000}), using data from the MACHO survey, showed that the
red giant variables form four roughly parallel sequences in a
period-magnitude diagram.  Three of these sequences could be
associated with fundamental, first and second overtone pulsation. The
explanation of the fourth sequence is not clear yet (Wood
\cite{Wood2000}, Hinkle et al.~\cite{Hinkle2002}). Recently, Cioni et
al.~(\cite{Cioni2001}) presented a survey of variable red giants in
the LMC based on data from the EROS-2 microlensing survey (Lasserre et
al. \cite{Lasserre2000}).
The work of Cioni et al.~focused on the logP$-$K relation confirming
three of the relations found by Wood, and they discussed the behaviour
of different groups of variables in near infrared colour-magnitude
diagrams.

The present paper relies on the variability information contained in the
AGAPEROS variable star catalogue (Melchior et
al. \cite{Melchior2000}). Here, we extend and analyse the corresponding light curves, relying
on the EROS-1 microlensing survey data set (Ansari et al. \cite{Ansari95}, Aubourg et
al. \cite{Aubourg95}). These data have been obtained between December
1991 and April 1994. The second half of the data set therefore
overlaps with the MACHO survey. We combine the EROS data with IJ$\rm
K_{S}$ photometry of the DENIS survey (Epchtein et
al. \cite{Epchtein97}), with an approach similar to the work of Cioni
et al. \cite{Cioni2001}. The intention of our work is to discuss the
light change of red variables on a large and homogeneous sample and to
compare the results for LMC red variables with the corresponding
objects in the Galactic Disk and Bulge.

\section{Data}
We have studied the AGAPEROS catalogue of 584 variable stars detected
over a 0.25 deg$^2$ field in the LMC Bar (Melchior et al. 2000,
hereafter referred to as Paper I). These stars have been selected on the
basis of their variability on a 120-days window with a bias towards
long-timescale variations ($>$ few days), which are not necessarily
periodic. The original data set has been taken at ESO by the EROS-1
collaboration (Arnaud et al., 1994a,b, Renault et al. 1997) using a
40cm telescope, equipped with a wide field camera composed of 16 CCD
chips, each of 400$\times$579 pixels of 1.21 arcsec (Arnaud et
al. \cite{Arnaud94b}). We use 9 chips, and study light curves in the
red ($\bar{\lambda}=670$nm) filter. Table~\ref{dataset} gives a short
description of the dataset. See Melchior et
al. (\cite{Melchior98},\cite{Melchior99}) for a more detailed
description of the data treatment.

The study of the position of these stars in the colour-magnitude
diagram showed that this catalogue is dominated by a population of
Long Timescale \& Long Period variables, while a few ``bluer''
variables have also been detected. A cross-correlation with various
existing catalogues showed that about 90$\%$ of those variable objects
were undetected before.  We extended the corresponding light curves to
the whole EROS-1 database of the LMC (900-days). We improved the
photometry of the corresponding light curves with image subtraction
using the ISIS2.1 algorithm of Alard (\cite{Alard2000}).
We follow the same definition of the magnitude system as in Paper I,
but we rely here on image subtraction photometry.

\begin{table}[h]
\caption{Description of the dataset}
\begin{flushleft}
\begin{tabular}{lr}
\hline \noalign{\smallskip}
time range (JD) & 2448611 - 2449462\phantom{*}\\
seasonal gaps & 2448721 - 2448860\phantom{*}\\
	& 2449076 - 2449207$^{*}$\\
mean number of data & \\
in each light curve & 448\phantom{*}\\
mean sampling & 1.6 days\phantom{*}\\
\hline \noalign{\smallskip}
\end{tabular}
\end{flushleft}
\begin{flushleft}
$^{*}$ 54 light curves have a larger gap of 482 days.
\end{flushleft}
\label{dataset}
\end{table}

\begin{figure*}[t]
\resizebox{\hsize}{!}{\includegraphics{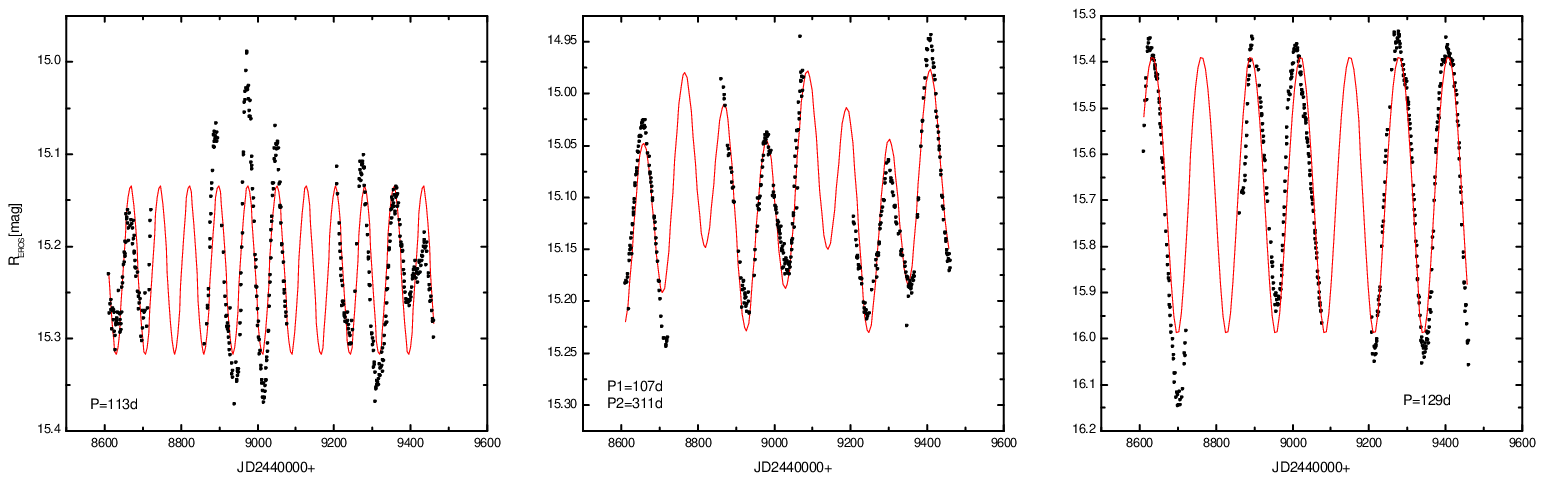}}
\resizebox{\hsize}{!}{\includegraphics{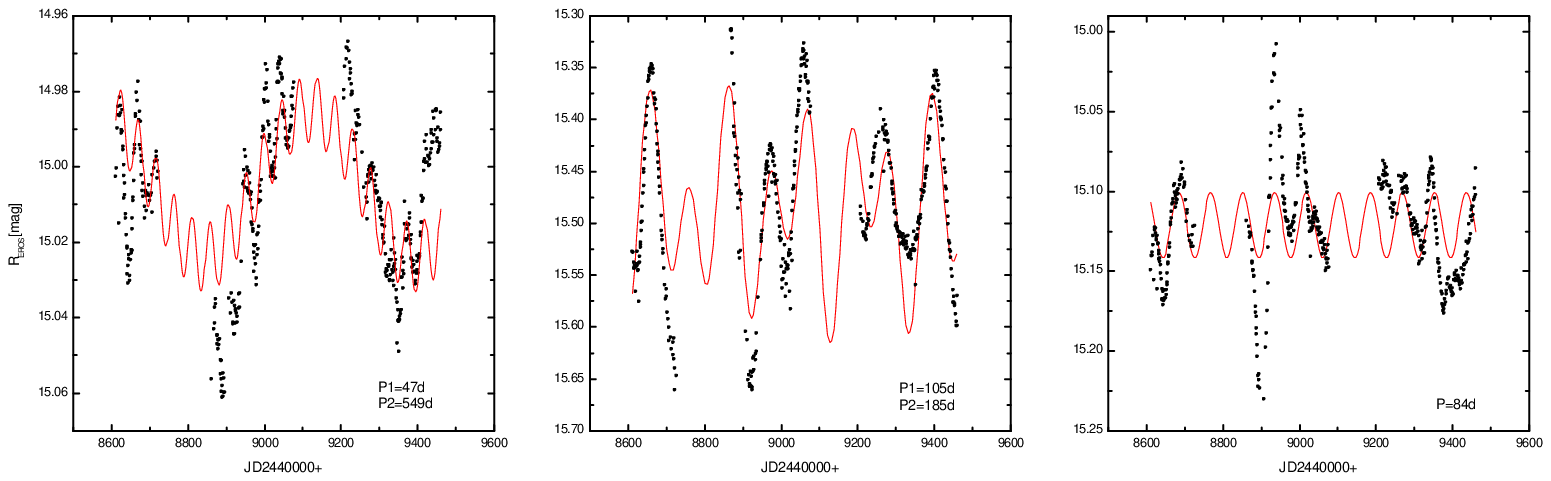}}
\resizebox{\hsize}{!}{\includegraphics{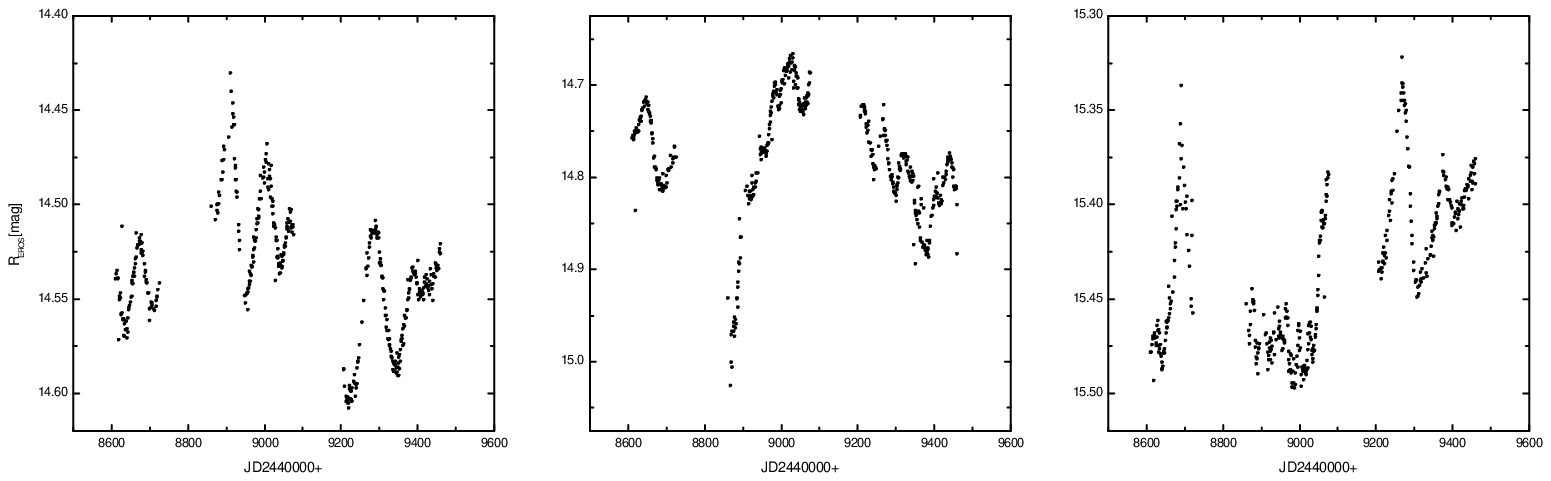}}
\caption{Example light curves for each group of variables: The upper
row illustrates regular light variation, the middle panels shows
representatives of the semiregular variables, and the bottom
row gives light variations classified as irregular. Periods
used for the fit are given in the plot. The amplitudes ($\rm \Delta R_{\rm EROS}$)
are given in mmag}
\label{sample}
\end{figure*}

The published DENIS catalogue (I, J and $\rm K_{\rm S}$) for the LMC
(Cioni et al. \cite{Cioni2000}) has been used to make a
cross-identification between the AGAPEROS variables and the DENIS
magnitudes. A search radius of 3\arcsec was chosen to avoid
misidentification.  Out of the 584 variables 468 were detected by
DENIS ($\sim$ 80\%). Note that the positional accuracy of these
variables is about 1\arcsec, as discussed in Paper I.


\section{Classification of the variables}
Classically, three types of variable red giants have been defined
(General Catalogue of Variable Stars, GCVS, Kholopov et
al.\,\cite{GCVS}): Mira-type variables show periodic large amplitude
variations with time scales typically of the order of 200 to 500
days. Semiregular variables (SRVs) show a less regular behaviour and a
smaller amplitude.  A typical time scale of the variation can be
found, but the light curve shows phases of irregularity as well. The
GCVS has introduced a limiting amplitude of 2.5 mag to separate Miras
and SRVs. While even within the GCVS this rule has not been strictly
applied (see e.g. the SRV W Hya), several investigators used this
simple criterion for classification (e.g.~Alard et al.~2001, Cioni et
al.~\cite{Cioni2001}). The artificial nature of this division has been
criticized already e.g.~by Kerschbaum (\cite{Kerschbaum93}).  The
third group of variables are the irregular variables. It is still not
clear if such stars really exist or if these objects are simply not
observed well enough to detect the same amount of periodicity as in
the SRVs (e.g.~Lebzelter et al.~\cite{LKH95}).

In this work, we used a different approach to classify the light curves
of the red giants in our sample. The classic classification system
depends on whether a more-or-less constant period can be found and also depends
on an arbitrary amplitude limit. Here we adopt a new scheme which is based
on how well the light curve can be described by one or two periods only and
where amplitude plays no role. In this way we are able to separate the
two effects amplitude and regularity.

We based the classification on the
regularity of the light curve on a visual comparison of the
light change with a combination of up to three sine curves.  To derive the periods, a
Fourier analysis of the light curves (based on the program Period98 by
Sperl (\cite{Sperl98})) has been applied.  Semiregular and irregular
light changes result in a large number of peaks of similar strength in
the Fourier spectrum (Lebzelter \cite{Lebzelter99}, see
below). 
Therefore the periods used for the fit have been selected from peaks in the 
periodogram by visual inspection. The amplitudes of the peaks were the 
starting point for the selection of the periods. Naturally, this selection 
is influenced by aliases. Fig.2 shows the typical spectral window of our 
data that has been used to identify spurious peaks. The selected periods 
were always cross checked by a visual comparison wiht the light curve. For 
unclear cases a second Fourier analysis was made with the primary period 
subtracted.

As
a first approach to this large amount of light curve data we made no
attempt to fit every detail of the light curves but identified the
major period(s) to roughly resemble the overall light change. A more
detailed fitting, as it was done by e.g.~Kerschbaum et
al.~(\cite{Kerschbaum2001}) for a small number of SRVs in the solar
neighborhood, is planned.  Finally, we stress that the total available
baseline of the data set did not allow to derive periodicities on time
scales longer than 900 days.  The classification is based on three
years of observation and represents the behaviour of each object over
the 900-days window.  Stars classified as {\it irregular} may show
some periodicity on a longer time scale or during a different time
interval. The amplitudes were estimated visually from the lightcurve.

We classified the light curves
on the regularity and type of their light change into four groups:
\begin{itemize}
\item {\it Regular:} A constant cycle length is observed for the
available measurements. Some of these objects show some amplitude
variations or bumps in their light curves. This group includes also
stars which show a second period, if the two periods allow a very good
fit of the light curve. While this group will include almost all
objects that classically would have been classified as Miras, possible
small amplitude regular variables will be found in this class as
well. We therefore did not use the name "Miras" for this group. The
only regular pulsators we did not include here were cepheids and
obvious binaries, which have been classified as {\it Other}.
\item {\it Semiregular:} The cycle length is variable, but
some kind of periodicity is visible. The stars show
up to three strong peaks in the Fourier spectrum. In some cases the
light change can be fitted rather well with three or four periods.
\item {\it Irregular:} These stars do not show any significant
periodicity in their light change. Their light change
occurs on time scales typical for AGB stars (i.e.~a few 10 to
a few 100 days). Typically, the Fourier spectrum
shows a large number of peaks with similar strength. 
\item {\it Other:} This group includes stars with a large fraction of
bad data points, stars that turned out to be constant (misidentification
due to some erroneous data points), and stars with a luminosity variation
atypical for long period variables (including a number of
binaries). A large fraction of these objects have also no DENIS data.
\end{itemize}

\begin{figure}
\resizebox{\hsize}{!}{\includegraphics{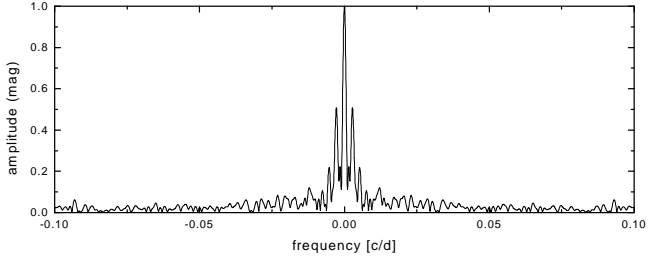}}
\caption{Spectral window for the light curves used in this paper.}
\label{window}
\end{figure}

Fig.~\ref{sample} shows a sample of regular, semiregular and
irregular light curves.  Note that our classification does not take
into account the amplitude of the variation as in the GCVS
classification. The examples were selected to represent the different
expressions of variability found in the three groups. Among the {\it
regular} variables, we included examples of amplitude variations (top
left in Fig.~\ref{sample}), variables with two periods (top middle)
and classical Mira variables (top right).

\begin{figure}
\resizebox{\hsize}{!}{\includegraphics{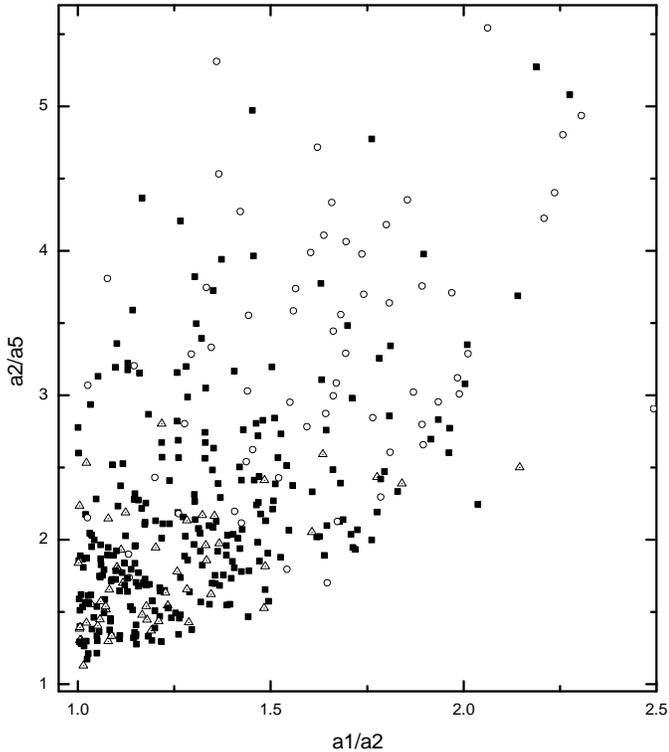}}
\caption{Ratio of the two strongest peaks of the Fourier amplitude spectrum
versus the ratio of the second and fifth strongest peak. Open circles denote
regular variables, filled boxes indicate semiregular variables and open triangles
mark irregular variables. A few objects found at even higher ratios are not included
in the plot.}
\label{classif}
\end{figure}

\begin{figure}
\resizebox{\hsize}{!}{\includegraphics{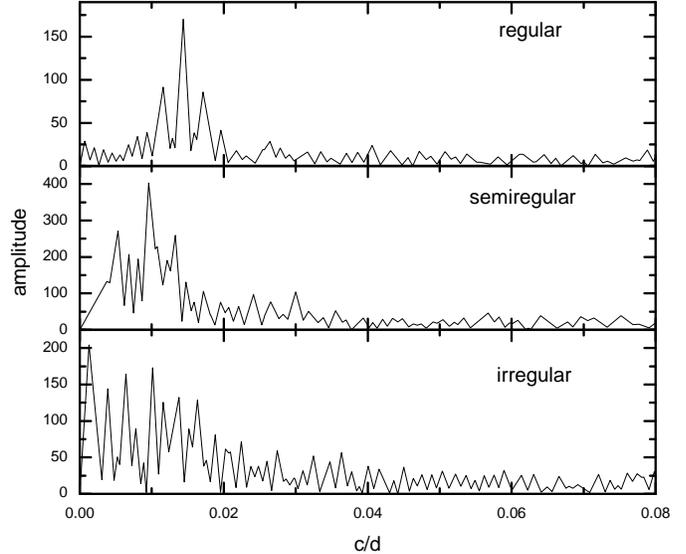}}
\caption{Typical Fourier amplitude spectra for a regular, a semiregular and
an irregular variable, respectively.}
\label{fourspe}
\end{figure}

Naturally, this classification remains somewhat subjective.  However,
we made an attempt to check the homogeneity of our classification by
using the Fourier spectra of the light curves. In Fig.~\ref{classif},
we plot the amplitude ratio of the strongest and the second strongest
peak against the ratio of the second and the fifth strongest peak. The
advantage of our sample is that all light curves have a similar
sampling and therefore a similar spectral window. Examples for Fourier
spectra and a spectral window are given in Fig.~\ref{fourspe} and
\ref{window}, respectively.  A small number of stars has been excluded
from this plot as their time coverage is not as good as for the
majority of the sample.

As mentioned above, a semiregular or irregular light curve typically
results in a number of peaks of similar strength in the Fourier
spectrum. Stars classified as {\it regular} have only one or two
strong peaks in their Fourier spectrum. They should therefore be found
on the right-hand side and the top side of Fig.\,\ref{classif}. Stars
with a single period are on the right, stars with a second period in
the upper left region of the plot. Note that there is no correction for
aliases in this approach. From the spectral window (Fig.\,\ref{window})
one would expect to find stars with a single period at a ratio a1/a2 of
about 2. 

For the {\it regular} variables the fifth strongest
peak is typically already at the noise level and was used as a reference
point. Note that we did not use more than three periods for each object in
the following analysis. On the other hand, {\it
irregular} variables should be found in the lower left corner of the
plot. {\it Semiregular} stars are expected in between.
Fig.\,\ref{classif} shows this classification indicated by
different symbols.  

We observe that our classification criteria is coherent within our
sample. However, for an individual object, Fig.\,\ref{classif} is not
usable for classification as the borders between the three classes are
not well defined.

For each star classified as {\it regular}, {\it semiregular} or {\it
irregular} a typical amplitude of the light variation was
determined. In the case of semiregular and irregular variables the
light amplitude can change dramatically.  In these cases, we used a
mean value of the variation. As no standard Johnson filters have been
used a direct comparison of the amplitude values found here and those
given in the GCVS or the MACHO catalogue is not possible.

\section{Comparison with results from the MACHO survey}
We searched the MACHO Variable Star Catalogue
(http://wwwmacho.mcmaster.ca/Data/MachoData.html) for variables within
the fields covered by our sample and classified as LPV.WoodA,
LPV.WoodB, LPV.WoodC and LPV.WoodD, respectively.  The catalogue
released on the web includes only a subsample of all variables found
in the MACHO survey. It was therefore not surprising that we did not
find all stars of our sample in the released MACHO catalogue, which is
probably not complete in the area we are concerned with. In total, we
found 36 MACHO LPVs that are within the fields we investigated. 25 of
them had counterparts in our sample\footnote{In the course of this
comparison, we found that the star 78.5861.10/80.6466.5194 has two
entries in the web based MACHO catalogue}.  We assume that the
remaining 11 stars are located on defects or borders of the CCD chips,
but we did not investigate these objects further.

We applied the same analysis to the red MACHO light curves of these
25 stars. Results are listed in Table\,\ref{compMACHO}.
The same classification was reached for 21 stars in our
comparison, in two further cases either the MACHO or the AGAPEROS
light curve was not of sufficient quality for the analysis.
Interestingly, the two remaining objects with different classifications
both show a higher degree of regularity in the MACHO data. In these
cases our dataset was obviously not covering enough light cycles to
reveal the regularity. We can estimate from this result that for less than
10\% of the AGAPEROS light curves the regularity was not detected correctly.

22 of the LPVs in common with the MACHO catalogue agree in the main period
within a few percent. The values we derived for the MACHO light curves are
in good agreement with the results listed in the MACHO catalogue. However, the
catalogue gives only one period for
each object, so in cases we found multiple periods in our sample stars
only one of them could be compared.  Several of the secondary periods
found in the AGAPEROS stars also agree quite well with values from the 
MACHO data. Very long periods could not be detected with the shorter
time series of AGAPEROS data. The differences in some secondary periods
illustrates the difficulty to derive a unique fit of these light
curves with more than one period (see also Kerschbaum et al.\,\cite{Kerschbaum2001}).

\begin{table*}
\caption{Comparison of MACHO and EROS data}
\begin{flushleft}
\label{compMACHO}
\begin{tabular}{lrrc|rrc}
\hline \noalign{\smallskip}
\multicolumn{4}{c}{MACHO} & \multicolumn{3}{c}{AGAPEROS}\\
F.T.S. number & Period 1 & Period 2 & Classif. & Period 1 & Period 2 & Classif.\\
\hline \noalign{\smallskip}
77.7549.37   & (74 d)&        & bad data$^{*}$   & 77 d & & semireg.\\
77.7550.65   & 593 d &   77 d & regular  &       &        & irreg.\\
77.7671.284  & 343 d &        & semireg. & 346 d &   39 d & semireg.\\
78.5616.19   &  68 d &   73 d & semireg. &  67 d &   82 d & semireg.\\
78.5737.16   & 120 d &        & regular  & 119 d &        & regular\\
78.5737.19   & 346 d & 3884 d & regular  & 340 d &        & regular\\
78.5739.75   &  96 d &        & regular  &  98 d &        & regular\\
78.5861.76   & 287 d &  160 d & semireg. &       &        & bad data\\
78.5981.182  & 193 d &        & regular  & 189 d &        & regular\\
78.5978.71   &       &        & irreg.   &       &        & irreg.\\
78.6099.145  & 128 d &        & regular  & 129 d &        & regular\\
78.6223.71   & 352 d &        & semireg. & 338 d &   52 d & semireg.\\
78.6343.57   & 128 d &        & regular  & 128 d &        & regular\\
78.6345.14   & 239 d &  126 d & semireg. & 225 d &  120 d & semireg.\\
78.6345.30   & 130 d &        & regular  & 129 d &  236 d & regular\\
78.6461.2171 & 437 d &   58 d & semireg. & 454 d &   51 d & semireg.\\
78.6466.18   & 338 d &        & regular  & 327 d &        & regular\\
78.6583.23   & 338 d &        & semireg. & 345 d &   56 d & semireg.\\
78.6586.61   & 121 d &        & regular  & 125 d &   67 d & semireg.\\
78.6707.35   &  89 d &        & regular  &  88 d &        & regular\\
78.6824.2327 & 150 d &        & semireg. & 150 d &  315 d & semireg.\\
78.6826.70   &  86 d &        & regular  &  86 d &        & regular\\
79.5863.25   & 91 d  & 1143 d & semireg. &  92 d &  274 d & semireg.\\
\hline \noalign{\smallskip}
\end{tabular}
\end{flushleft}
\begin{flushleft}
$^{*}$ The blue MACHO data give a period of 74 days.
\end{flushleft}
\end{table*}

The good agreement in the classification between the AGAPEROS and the MACHO
data, computed over independent data sets (1991-1994 for AGAPEROS,
1992-2000 for MACHO), strengthens the validity of our approach.

\section{General characteristics of the sample variables}
{\it Semiregular} variables clearly dominate our sample of late type
giants. 583 light curves have been analysed in total. 112 of them have
been classified as {\it other}. Among the remaining 471 objects we
classified 18\,\% as {\it regular}, 67\,\% as {\it semiregular} and
15\,\% as {\it irregular}. Due to different classification criteria a
comparison of this result with other investigations is difficult.
Cioni et al.\,(\cite{Cioni2001}) found that 65\,\% of the AGB stars
are variable. As in our sample semiregular variables (in their case
small amplitude red variables) are the dominant group of objects, only
12\,\% of their stars have been classified as Miras.

\begin{figure}
\resizebox{\hsize}{!}{\includegraphics{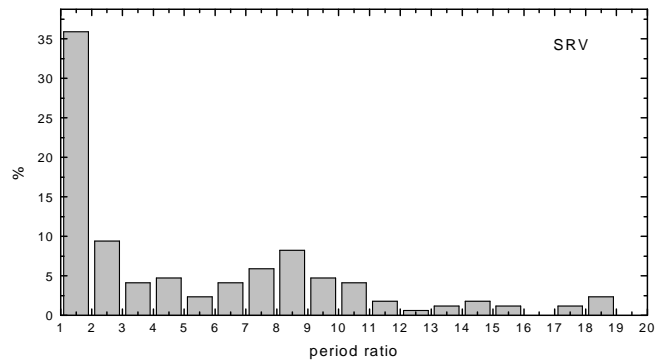}}
\caption{Ratio between the short and the long period of semiregular
variables with two periods.}
\label{agapratio}
\end{figure}

If more than one period is detected, we have chosen a 'primary' period
based on its amplitude.
However, this decision is somewhat subjective when
the amplitudes are similar. ``Second periods'' are present in at
least 54\,\% of the stars classified as {\it semiregular} and
about 18\,\% of the {\it regular} variables. For the {\it semiregular}
variables the ratio between the short and long period
is between 1 and 2 for about 36\,\% of the objects (Fig.\,\ref{agapratio}),
the other {\it semiregulars} have period ratios between 2 and 15. This result is similar
to what has been found in previous investigations of SRVs in
our Galaxy (e.g.\, Kiss \& Szatmary \cite{Kiss2000}).
Among the {\it regular} variables with two periods about one third of the stars
show a period ratio below 2.

\begin{figure}
\resizebox{\hsize}{!}{\includegraphics{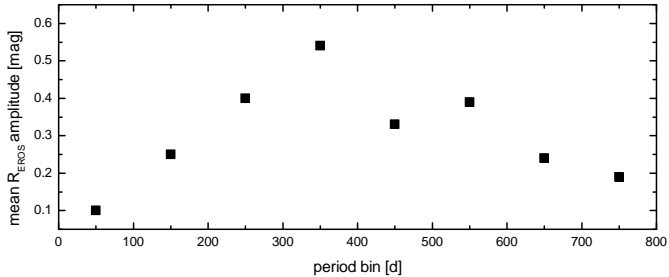}}
\caption{Mean amplitude for each period bin. Only the primary period has
been used in this plot.}
\label{periampl}
\end{figure}

\begin{figure}
\resizebox{\hsize}{!}{\includegraphics{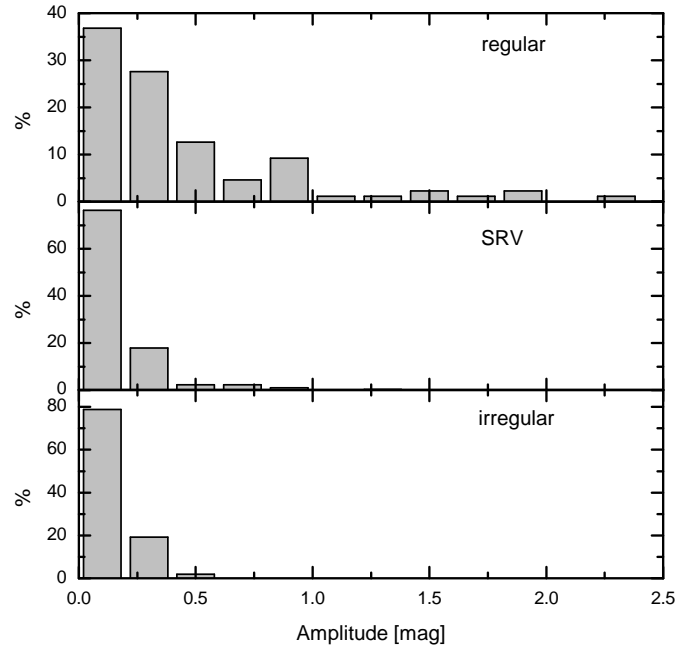}}
\caption{R$_{\rm EROS}$ amplitude distribution for regular, semiregular and
irregular variables, respectively. As discussed in Sect. 3, the
amplitudes are estimated visually from the light curves.}
\label{ampldist}
\end{figure}

Comparing amplitude with period shows a maximum amplitude for periods
between 300 and 400 days, given our 900-day window. No large amplitude
variables with periods below 100 days have been found. Details are
illustrated in Fig.\,\ref{periampl} where the mean amplitude has been
calculated for each period bin.  Alard et al.\,(\cite{Alard2001})
found a similar increase of mean amplitude from small to long periods
in a sample of AGB variables in the Galactic Bulge. However, their
sample shows a maximum amplitude at periods around 250 days, as they excluded
all miras from their analysis.
Fig.~\ref{ampldist} summarizes the amplitude distributions within each
class of variables. All three classes are dominated by small amplitude
objects.  Large amplitude stars occur exclusively among {\it regular}
variables, all {\it irregular} objects are small amplitude stars.

\begin{figure}
\resizebox{\hsize}{!}{\includegraphics{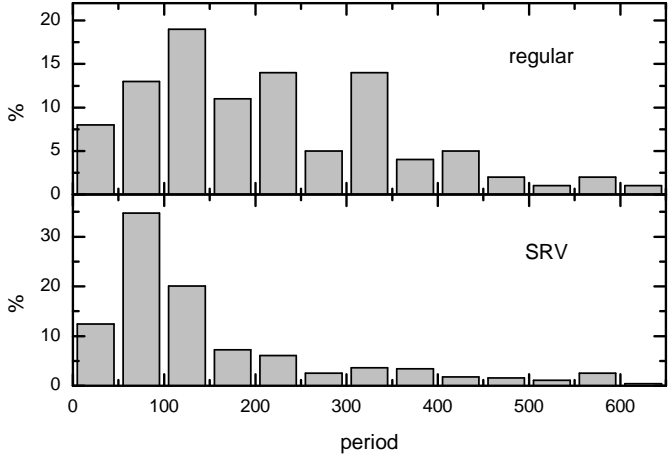}}
\caption{Period distribution of regular and semiregular AGAPEROS variables.
If a star has two periods, both have been included. Periods are given
in days. The period determination is based on Fourier analysis as
discussed in Sect. 3.}
\label{Perioddistribution}
\end{figure}

We will now mainly concentrate on regular and semiregular variables.
Fig.~\ref{Perioddistribution} shows the period distribution of the
AGAPEROS sample. Both groups of objects (regular and semiregular) show
a maximum at the shortest periods: Therefore, the {\it regular}
variables cannot be simply related to the class of Mira variables. No
Galactic Mira with a period below 100 days is known.  Furthermore,
Miras typically show large amplitude variability while the short
period {\it regular} stars in our sample all have small amplitudes.
The "classical" Miras probably form the second maximum in period
distribution of the {\it regular} variables around 350 days, similar
to the Galactic Miras. We conclude that the group classically known as 
semiregular variables
may contain a substantial number of periodic variables. On the other hand
it is known from long time series of Galactic semiregular variables that
these stars can show phases of periodic behavior, so that these stars
may exhibit semiregular behavior on longer time scales.

\section{Near-Infared data}
\begin{figure}
\epsfysize=10.0cm

\centerline {\epsfbox[10 10 570 750]{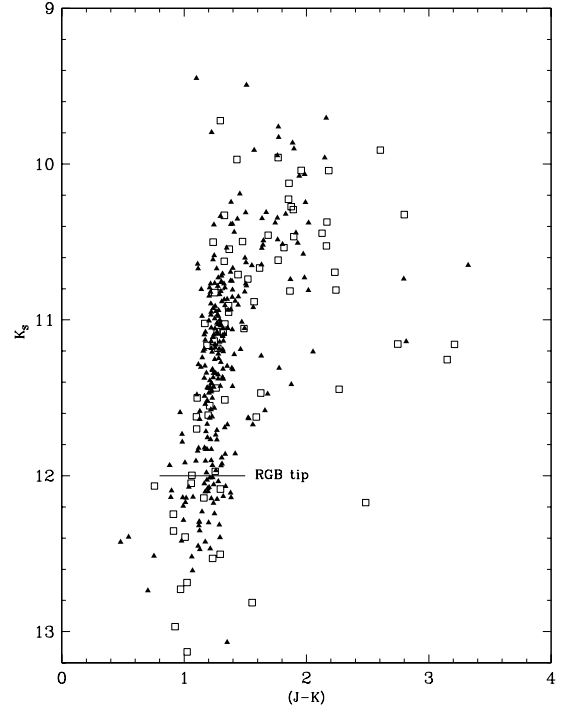}}
\caption{Colour-magnitude diagram for DENIS/AGAPEROS stars. Regular variables
are indicated by open squares, semiregular variables  by filled
triangles. The horizontal line indicates the tip of the red giant
branch (RGB).}
\label{CMD}
\end{figure}

The near-infrared data from the DENIS survey allow to characterize the
variables of our sample in more detail concerning their luminosity and
chemical composition.  Fig.~\ref{CMD} shows the $\rm
K_{S}/(J-K_{S})$ diagram for the AGAPEROS variables.  One can clearly
see that the majority of the sources are located above the tip of the
Red Giant Branch (hereafter RGB-tip) which is for the LMC about
12.0\,mag in $\rm K_{S}$ (Cioni et al. \cite{Cioni2000a}).
We find regular and semiregular variables which are below the
RGB-tip. These objects have rather short periods ($<$\,100\,days) and
could be AGB stars in the early evolutionary phase (early-AGB phase)
or variable stars on the red giant branch.

\begin{figure}
\epsfysize=10.0cm

\centerline {\epsfbox[10 10 570 750]{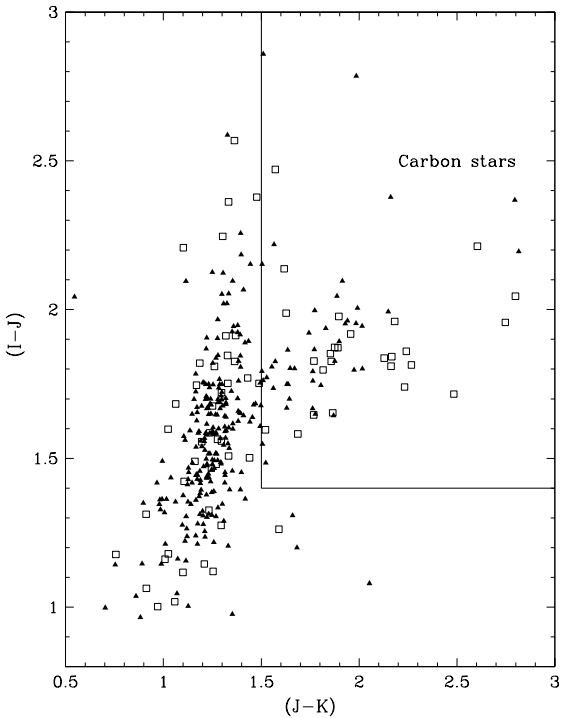}}
\caption{DENIS colour-colour diagram for AGAPEROS variables. The box
indicates the approximate location of carbon-rich objects (see Loup et
al. 2002). Regular variables and semiregular variables are indicated
by open squares and filled triangles, respectively.}
\label{CCD}
\end{figure}

Carbon-rich objects are characterized by their red (J--K) and (I--J) colour
compared to the oxygen-rich sequence (see Cioni et
al. \cite{Cioni99}). However, as noted by Loup et
al. (\cite{Loup2002}) the colour-colour diagram is just a statistical
tool to distinguish between oxygen-rich and carbon-rich objects.
  Fig.~\ref{CCD} shows the $\rm (I-J)_{0}$ vs $\rm
(J-K)_{0}$ diagram.  Obviously, the ratio of regular to semiregular
variables is smaller for the oxygen-rich stars than for the
carbon-rich objects. This suggests that the majority of the
semiregular variables are less massive than Miras which prevents them
from becoming carbon stars. 

We do not find any significant difference in colours or luminosites
 between SRVs with one single period and  SRVs with multiple periods.

\begin{figure*}[t]
\epsfxsize=8.5cm

\centerline {\epsfbox[101 137 630 470]{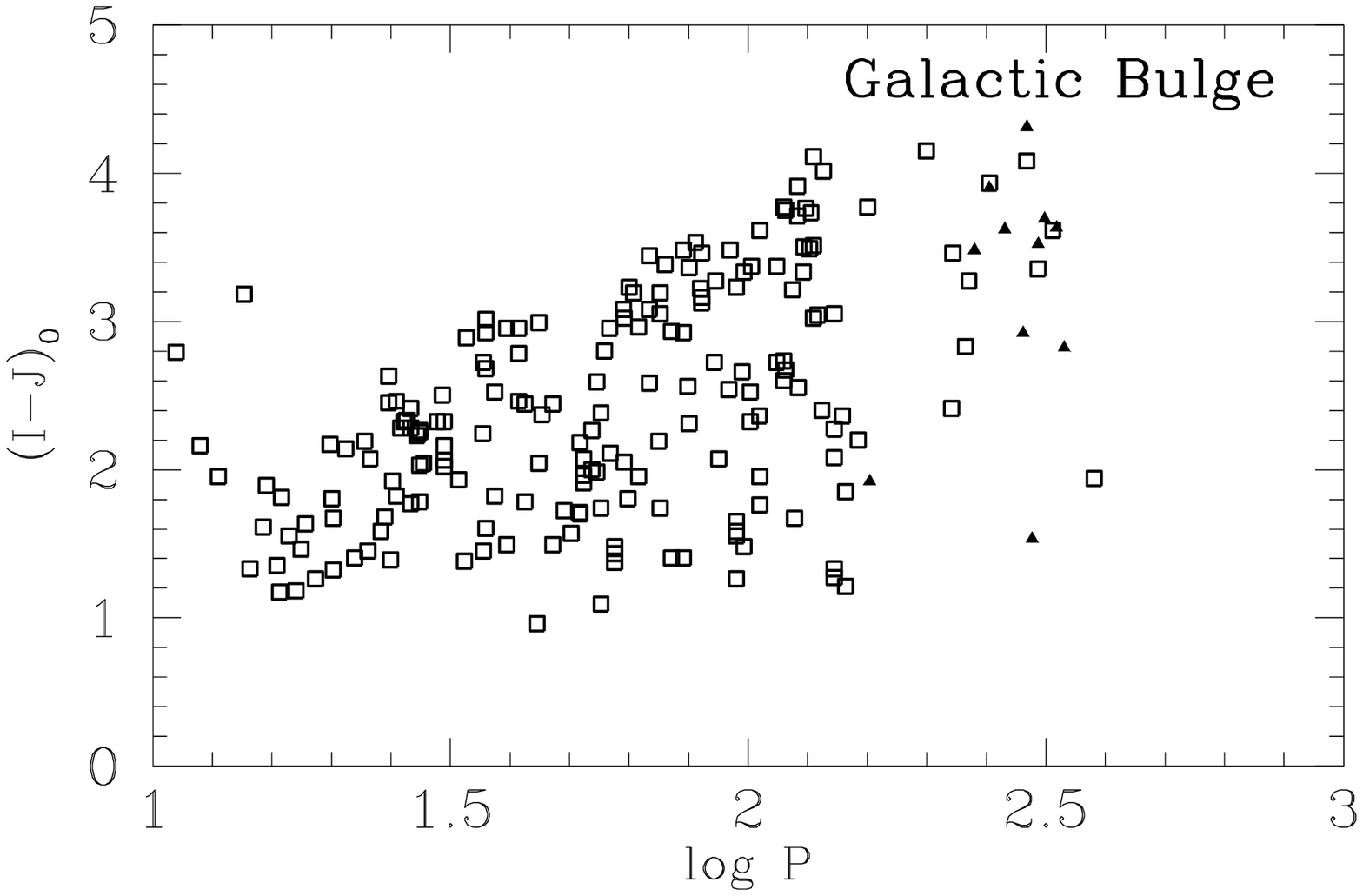} \epsfxsize=8.5cm 
 \epsfbox[101 137 630 470]{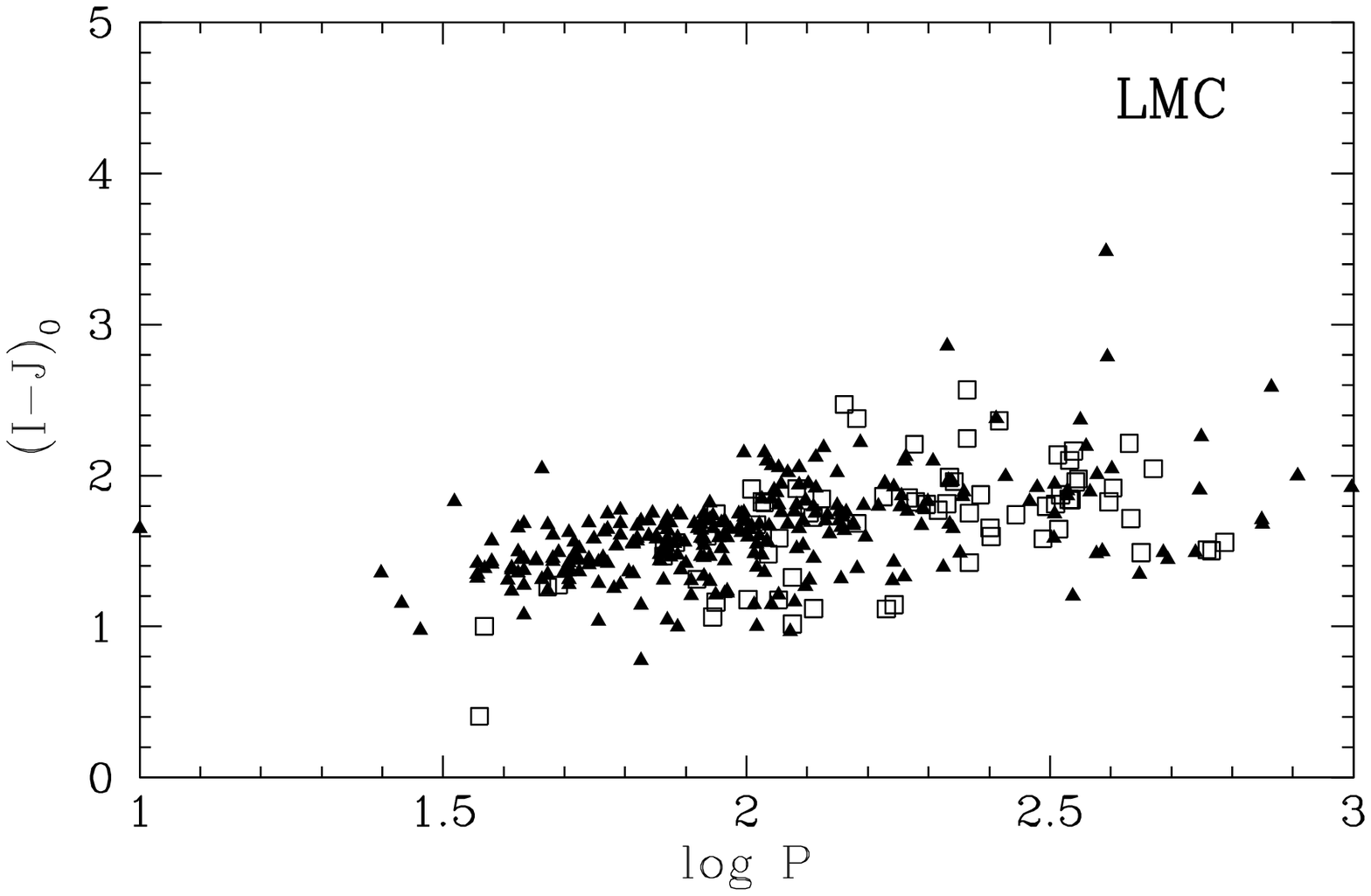}}
\caption{LogP vs (I--J) relation for MACHO variables in Baade's window
(Schultheis \& Glass \cite{Schultheis2001}) compared to AGAPEROS
variables in the LMC.  The open squares on the left panel indicate the
SRVS, while the filled triangles the Mira variables.  On the right
panel, same symbols as in Fig.~10. The periods are given in days.}
\label{logPIJ}
\end{figure*}

\subsection{Colour-Period diagrams}

For the LMC bar, it is obvious from Fig.~\ref{logPIJ} that the
AGAPEROS variables follow a tight logP vs I--J relation.  It is
important to emphasize that the I magnitudes of DENIS correspond to a
{\em{single}} epoch measurement and thus the logP vs. I--J diagram is
affected by the scatter due to the amplitude variation of each source.

Relying on MACHO data in the Galactic Bulge, Schultheis \& Glass
(\cite{Schultheis2001}) demonstrated that semiregular variables in the
Galactic Bulge show a noticeable scatter in I--J (3--4\,mag) along the
logP vs I--J relation. The most significant difference between the
Galactic Bulge and the LMC is the smaller range in I--J for the LMC
($\sim$ 2\,mag) than for the Bulge ($\sim$ 4 mag). (see
Fig.\,\ref{logPIJ})

The I band for M stars is mostly affected by the strong TiO and VO
molecular absorption (Turnshek et al. \cite{Turnshek85}, Lancon \&
Wood \cite{Lancon2000}).  Schultheis et al. (\cite{Schultheis99})
showed that lower metallicity is correlated with weaker TiO band
intensities, corresponding to bluer I--J colours.  The large scatter
and the wide I--J range in the Galactic Bulge sample compared to the
LMC might be explained by the wide spread in metallicity compared to
the Magellanic Clouds.  However, the difference in the I--J range
between the Galactic Bulge ($\rm 1 < (I-J)_{0} < 5$) and the LMC ($\rm
1 < (I-J)_{0} < 3$) seems rather large. A more detailed quantitative
analysis, using realistic model atmospheres of AGB stars (including
metallic lines), is necessary to fully understand this systematic
difference in the I--J colour between the Galactic Bulge and the LMC.

\begin{figure}
\epsfxsize=8.5cm
\centerline {\epsfbox[20 30 740 560]{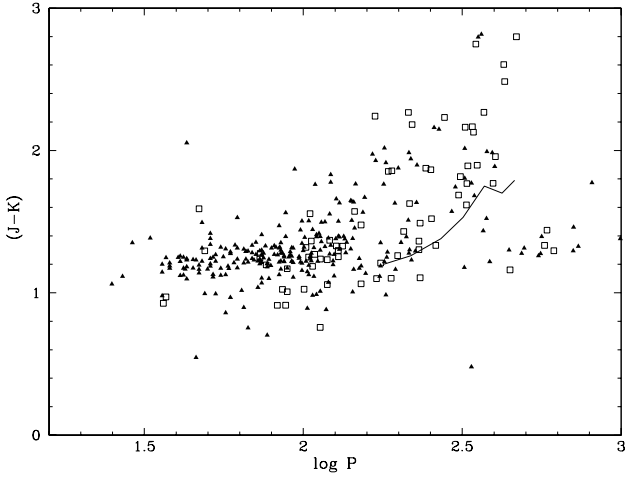}}
\caption{logP vs (J--K) relation for AGAPEROS variables in the LMC. The  line indicates  the average colours of SgrI Miras for various period groups (Glass et al. \cite{Glass95}). The symbols are the same as in Fig.~10}
\label{logPJK}
\end{figure}

Fig.~\ref{logPJK} displays the J--K colours of the AGAPEROS
variables as a function of their period.  The majority of the SRVs
appear to follow a different period-colour relation with a slope
flatter than the regular variables. For comparison, we indicated in
Fig\,\ref{logPJK} the averaged colours of oxygen-rich Miras for the
SgrI field (Glass et al. \cite{Glass95}). The majority of our
long-period Miras ($\rm log\,P > 250^{d}$) follow the location of the
oxygen-rich Miras in SgrI.  The carbon rich objects (J-K\,$>$\,1.6)
seem to form a parallel sequence to the oxygen-rich Miras,  while the
long-period SRVs ($\rm P > 300^{d}$) do show clearly another
period-colour relation. These stars are located on the sequence D in
Wood's diagram (see Wood et al. \cite{Wood99} and discussion below)
and are SRVs with multiple periods. A few long-period Miras also
follow this sequence.  However, the scatter in this diagram increases
for $\rm log\,P > 2.3$ due to the contribution of the circumstellar
dust shell arising from mass loss.  Schultheis et
al.~(\cite{Schultheis99}) and Schultheis \& Glass
(\cite{Schultheis2001}) obtain similar results for semiregular
variables in the Galactic Bulge (see their Fig.~8).

\subsection{$\rm K_{S}$ vs logP diagram}
In the Large Magellanic Cloud, the Miras and the SRVs seem to form
distinct parallel sequences C,B,A which have been identified by Wood
(\cite{Wood2000}) as pulsators in the fundamental, first and the next
two higher overtones, respectively.  Wood et al.~(\cite{Wood99})
showed by comparison of observed periods, luminosities and period
ratios with theoretical models, that Miras are radial fundamental mode
pulsators, while semiregular variables can be pulsating in the 1st,
2nd or 3rd overtone, or even the fundamental mode.
The pulsation mode derived by Whitelock \& Feast (\cite{WF00}) from
diameter measurements of Miras in the Milky Way suggests first
overtone pulsation for Miras.  However, observations of radial
velocity variations of Miras (e.g.~Hinkle et al.\,\cite{HHR}) clearly
favour fundamental mode pulsation (Bessell et al.\,\cite{BSW96}).

\begin{figure}
\epsfxsize=8.5cm
\resizebox{\hsize}{!}{\includegraphics{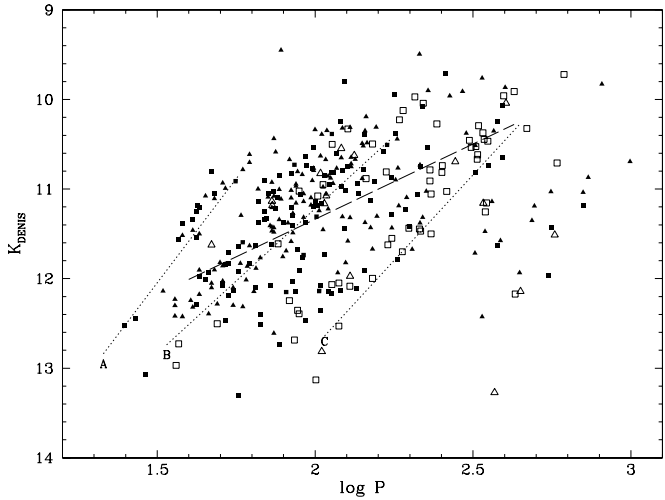}}
\caption{$\rm K_{S}$ vs logP diagram for AGAPEROS variables. The dashed line is the relationship suggested for local SRVs by 
Bedding \& Zijlstra (\cite{Bedding98}). The dotted lines labelled A,B
and C are eye fits to the sequence by Wood (\cite{Wood2000}).   
We use only the primary periods.  Open squares show regular variables with one single period 
while  regular variables with a second period are shown as open triangles. Semiregular variables 
with one single period are shown as filled squares while those with their second period are 
indicated as filled triangles.
Sequence D of Wood (\cite{Wood2000}) lies on
the right-hand side.}
\label{KlogP}
\end{figure}

In Fig.~\ref{KlogP}, we distinguish between semiregular variables with
one period and those having a second or even third pulsational period.
The location of our regular variables is consistent with the
PL-relation from Feast et al. (\cite{Feast82}) and Wood's sequence C
corresponding to fundamental mode pulsation. However, a few regular
variables are also found to be located on sequence B and A (first and
second overtones according to Wood (\cite{Wood2000})).
The majority of the SRVs follow Wood's sequence B although the scatter
is rather large ($\sim$ 0.5\,mag in $\rm K_{S}$ at a given period).
The SRVs situated on sequence A show very low amplitudes ($<$ 0.5\,mag
in $\rm R_{EROS}$) and typically no secondary periods.  While Cioni et
al.\,(\cite{Cioni2001}) found no objects on sequence A, we could
clearly confirm the existence of this PL-sequence. On sequence B and
C, we find both single periodic and multiperiodic objects. The
occurrence of single or multiple periodic behaviour does not depend on
the luminosity.

Several data points also mark sequence D of Wood
(\cite{Wood2000}). The large scatter in this part of the
K-logP-diagram is due to the limited time window of our data set. We
are therefore able to reproduce all four sequences found in the MACHO
data. The PL-relation for SRVs found by Bedding \& Zijlstra
(\cite{Bedding98}) from local objects could not be confirmed with our
data (see below).

\section{Discussion}
\subsection{Variability on the AGB}
\begin{figure}
\resizebox{\hsize}{!}{\includegraphics{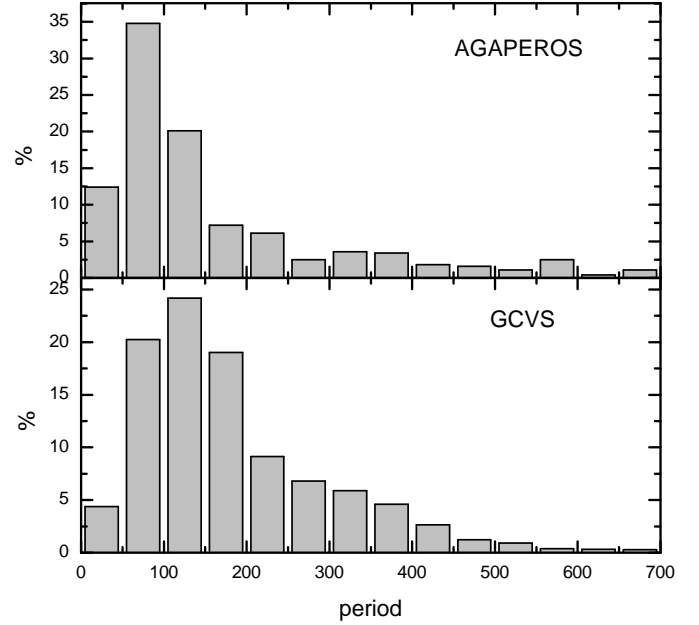}}
\caption{Period distribution of semiregular variables in our sample and in the GCVS.}
\label{agapgcvs}
\end{figure}

According to Fig.\,\ref{CMD} most of the variables in our sample are
on the AGB. Therefore, we can use our results to discuss the
variability during the AGB phase. Our classification system for the
type of variability aims to measure the regularity of the light
change. Even not taking into account variations in the amplitude of
the light change, we show that most stars have light curves that cannot
be fitted by the simple combination of one or two excited periods.
Regular variations are found with a wide range in period, while
semiregular variability typically occurs mainly on time scales below
150 days (see Fig.\,\ref{Perioddistribution}).  In
Fig.\,\ref{agapgcvs}, we compare the period distribution of the
semiregular variables in our sample with the Milky Way SRVs listed in
the GCVS. While in both cases the maximum of the distribution is at
short periods, the GCVS distribution shows a significantly larger
fraction of stars with periods longer than 150 days. These long
periods may have been missed by our rather short time window. 
One would also expect a bias of the GCVS sample towards large amplitude
variables as most of the data used there are based on photographic
measurements.
Furthermore, the period distribution from the GCVS given in
Fig.\,\ref{Perioddistribution} includes only one (main) period per
object, while for the AGAPEROS data we give also secondary periods
found for these stars.

\begin{figure}
\resizebox{\hsize}{!}{\includegraphics{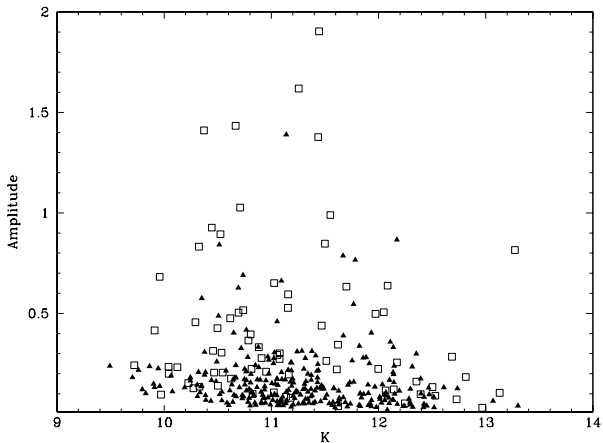}}
\caption{$\rm R_{EROS}$ amplitude versus $\rm K_S$. Open boxes denote {\it regular} variables,
filled triangles {\it semiregular} stars.}
\label{KsAmpl}
\end{figure}

Due to the separation of amplitude and regularity in our
classification system, we can explore the relation between these two
quantities. We find that large amplitude variation occurs almost
exclusively among the {\it regular} variables (see
Fig.\,\ref{ampldist}). However, there exist {\it regular} pulsators
with small amplitudes.  It is therefore not correct to classify all
red variables below a certain amplitude limit as semiregular.  A
division into large and small amplitude variables seems to be more
meaningful.  Large and small amplitude variables are both found all
along the AGB. This is illustrated in Fig.\,\ref{KsAmpl} where the
$\rm R_{EROS}$ light amplitude is plotted against the DENIS K band
measurement. Towards the tip of the AGB the fraction of regular as
well as large amplitude variables increases. Below the RGB-tip,
amplitudes become on the average smaller.  The occurrence of regular
and semiregular as well as small and large amplitude variables on the
AGB indicates that AGB stars have to be seen as a highly inhomogeneous
group. One reason for this may be a difference in stellar mass as
noted above. 

Summarizing, large amplitudes are well correlated with regular
pulsations, but we find no correlation between large amplitude and
stellar luminosity nor between small amplitude variability and
semiregularity of the light change.  This result is in agreement with
Wood et al.\,(\cite{Wood99}).

\subsection{PL-relation}
In the literature, the observed PL-relation of long-period variables
is considered to be the same in different environments such as the
LMC, the Galactic Bulge or globular clusters (see e.~g. Glass et
al. \cite{Glass95}, Feast et al. \cite{Feast2002}). It is therefore
independent of metallicity, contrary to the predictions of pulsation
theory (see e.~g. Wood\,\&\,Sebo \cite{Wood96}).  However, previous
studies were restricted to Mira variables mainly due to limitation of
sensitivity.  Thanks to the microlensing surveys such as EROS, MACHO
or OGLE, we can study {\em{systematically}} small amplitude variations
over a (still rather small) time interval. Wood (\cite{Wood2000})
found for the SRVs in the LMC different PL-relations for different
pulsational modes. However, these separations cannot be reproduced in
the Galactic Bulge (Schultheis \& Glass 2001). In addition, the
PL-relation of the solar neighborhood (Bedding
\& Zijlstra \cite{Bedding98}) looks different. 
Why is the PL-relation the same for Miras in different galactic
environments, but {\it not} for SRVs?

Fig.\,\ref{KlogP} shows the PL-relation for the AGAPEROS sample.  On
the one hand, the Mira variables, classically defined as long period
and large amplitude stars, concentrate along Wood's sequence C.  {\it
Regular} variables at shorter periods would not have been classified
as Miras. On the other hand, the semiregular variables (both according
to the classical and to our definition), are spread all over the
K-log\,P-plane.  Making one fit with all {\it semiregular} stars would
not result in a K vs.~log\,P relation.  In the solar neighborhood,
Bedding \& Zijlstra (\cite{Bedding98}) note that the SRVs are actually
found on two sequences: the first one corresponds to the LMC Mira
PL-relation (Wood's sequence C); the second one is located close to a
PL-relation derived from Galactic globular cluster LPVs shifted 0.8\,mag from the Whitelock globular cluster sequence (Whitelock
\cite{Whitelock86}), as shown in Fig.\,\ref{KlogP}. The Bedding \&
Zijlstra sequence, defined for SRVs, obviously mixes objects from
Wood's sequence B and C, as shown in Fig.\,\ref{KlogP}. The increase
towards longer periods is consistent with the larger fraction of long
period SRVs in the GCVS (Fig.\,\ref{agapgcvs}) assuming that the
detection of long periodic small amplitude variations is biased
towards bright objects. Therefore, three PL-sequences seem to be more
appropriate for semiregular variables. Multiperiodic stars are found
on all three sequences A, B and C (see Fig.\,\ref{KlogP}). Sequence D
is almost exclusively occupied by stars with two periods in agreement
with the suggestion from Wood (\cite{Wood2000}) that these long
periodic variations are either due to binarity or a pulsation mode
resulting from an interaction of pulsation and convection.  However,
there are also a few {\it regular} pulsating variables on this
sequence with only one period.  These stars would be definitely worth
further investigation.

Schultheis \& Glass (\cite{Schultheis2001}) showed that the
interpretation of the PL-relation of Bulge SRVs is rather complex due
to the depth of the Bulge ($\sim$ $\rm \pm 0.35^{mag}$, see Glass et
al. \cite{Glass95}) and the variable interstellar extinction. There is
no clear separation of the four sequences. We also showed that the LMC
variables are much more homogeneous in their metallicity than the
Bulge AGB stars (Fig.\,\ref{logPIJ}).  This would explain part of the
scatter in the K vs.~log\,P plot for the Bulge.

\subsection{Number densities}
The number of semiregular variables in comparison to the regular
variables is about a factor of 3.  If we use the selection criterion
of Cioni et al.\,(\cite{Cioni2001}), i.e.~all stars with $\rm R_{EROS}$
amplitudes smaller than 0.9 mag are SRVs, we end up with a ratio of
almost 37 between SRVs and Miras in our sample. This value is much higher than what
was found by Cioni et al.~($\sim$5), so we assume that our sample is
more complete at smaller amplitudes.  In the Galactic Bulge, Alard et
al. (\cite{Alard2000}) found that the proportion of SRVs with respect
to Miras is about a factor of 20. 
Most recently, Derue et al.\,(\cite{Derue02}) found a similarly large
ratio between semiregulars and miras in the Galactic spiral arms.
However, this ratio is of course
very sensitive to the classification of SRVs (see above).  For the
Galactic disk, Kerschbaum \& Hron (\cite{KH92}) found equal number
densities for Miras and semiregular variables. However, they note that
their sample of semiregular variables is probably not complete due to
the difficulties in detecting small amplitude variables.

Do we see in different environments the same ratio of SRVs to Mira
variables or does it depend on metallicity?  Vassiliadis \& Wood
(\cite{Vassiliadis93}) calculated lifetimes of the major evolutionary
phases for different initial masses and different metallicities. They
found that higher metallicity will increase the lifetime of the
early-AGB but decrease the lifetime on the TP-AGB. Miras stars
populate the TP-AGB, therefore in environments with higher
metallicities, such as the Galactic Bulge the lifetime of the TP-AGB
is shorter and thus the number densities should decrease. This might
explain the correlation between the ratio of SRVs to Miras and
metallicity. However, while a large fraction of our variables on the
TP-AGB are {\it regular} variables\footnote{In this case {\it regular}
variables and Miras can be assumed to be identical.} also {\it
semiregular} variables are found. Lebzelter \& Hron (\cite{LH99}) have
shown that for stars in the solar neighborhood stellar evolution goes
from SRVs to Miras. The {\it semiregular} stars found at a similar
luminosity as the Miras (see Fig.\,\ref{KlogP}) are therefore probably
not in the same evolutionary state or they have different
masses. Comparison of the number densities with expected lifetime is
therefore problematic.

A lower metallicity leads also to a shift of the AGB towards higher
temperatures in the HR diagram. The visual light change of these cool
variables is dominated by highly temperature sensitive molecules like
TiO (e.g.~Reid \& Goldston \cite{RG02}). If the stellar temperature is
higher, these molecules will play a minor role. Lower metallicity will
also make the TiO bands weaker. Therefore one would
expect that the visual amplitudes will in general be smaller for lower
metallicity. This would favour small amplitude variability in metal
poor environments and would explain the smaller fraction of large
amplitude objects in the LMC compared to the Bulge. It would also be
consistent with the complete lack of Miras in metal poor globular
clusters (Frogel \& Whitelock \cite{FW98}).

However, one has to be extremely careful concerning possible selection
effects, in particular for small amplitude variables. A homogeneous
survey of variable stars in different Galactic environments is
therefore needed.

\acknowledgements{
ALM thanks the EROS collaboration and in particular Jean-Baptiste
Marquette for his help with the light curves production with the image
subtraction method. ALM is extremely grateful to Claude Lamy who
performs the tremendous work of sorting the whole EROS-1 data set.  TL
has been supported by the Austrian Science Fund under project number
P14365-PHY. MS is supported by the Fonds zur F\"orderung der
wissenschaftlichen Forschung (FWF), Austria, under the project number
J1971-PHY. We wish to thank Josef Hron for fruitful discussion. Finally,
we wish to thank the referee for constructive comments. This
paper utilizes public domain data obtained by the MACHO Project,
jointly funded by the US Department of Energy through the University
of California, Lawrence Livermore National Laboratory under contract
No. W-7405-Eng-48, by the National Science Foundation through the
Center for Particle Astrophysics of the University of California under
cooperative agreement AST-8809616, and by the Mount Stromlo and Siding
Spring Observatory, part of the Australian National University.}

\end{document}